\documentclass[twocolumn]{aastex631}
\usepackage{amsmath}
\usepackage{multirow}
\usepackage{mathtools}
\usepackage[thinc]{esdiff}

\newcommand{\bz}{$\langle B_z \rangle$}

\received{----}
\revised{----}
\accepted{----}
\submitjournal{ApJ}

\received{May 31, 2023}
\revised{September 15, 2023}
\accepted{October 12, 2023}

\begin{document}

\title{Unstable phenomena in stable magnetospheres: searching for radio flares from magnetic OBA stars using VCSS}

\shorttitle{Flares from massive stars}
\shortauthors{Polisensky et al.}

\correspondingauthor{Emil Polisensky}
\email{emil.polisensky@nrl.navy.mil}

\author[0000-0003-3272-9237]{Emil Polisensky}
\affiliation{Code 7213, U.S. Naval Research Laboratory, 4555 Overlook Ave SW, Washington, DC 20375, USA}

\author[0000-0001-8704-1822]{Barnali Das}
\affil{Department of Physics and Astronomy, Bartol Research Institute, University of Delaware, 217 Sharp Lab, Newark, DE 19716, USA}
\affil{CSIRO Space and Astronomy, PO Box 1130, Bentley WA 6102, Australia}

\author[0000-0002-5187-7107]{Wendy Peters}
\affiliation{Code 7213, U.S. Naval Research Laboratory, 4555 Overlook Ave SW, Washington, DC 20375, USA}

\author{Matt E. Shultz}
\affil{Department of Physics and Astronomy, Bartol Research Institute, University of Delaware, 217 Sharp Lab, Newark, DE 19716, USA}

\author[0000-0002-1912-1342]{Eugene Semenko}
\affiliation{National Astronomical Research Institute of Thailand, Chiangmai 50180, Thailand}

\author[0000-0001-6812-7938]{Tracy E. Clarke}
\affiliation{Code 7213, U.S. Naval Research Laboratory, 4555 Overlook Ave SW, Washington, DC 20375, USA}



\begin{abstract}

Although the majority of hot magnetic stars have extremely stable, $\sim$kG strength surface magnetic fields with simple topologies, a subset undergo small-scale explosions due to centrifugal breakout (CBO). The resulting small-scale flares are typically below the sensitivity of current magentospheric diagnostics and do not generate detectable transient signatures.  However, a recently reported radio flare from the hot magnetic star CU Vir suggests that some of the most energetic events do reach detectable levels. Motivated by this, we searched for transient radio sources in the first two epochs of the VLITE Commensal Sky Survey (VCSS) at the position of 761 hot magnetic stars. We report three detections. A false association analysis shows a less than 1\% probability that the sources are imaging artifacts. We then examine the stellar parameters of the three stars to understand if they are likely to produce flares. We conclude that while at this stage we cannot make a definitive association of the detections with the stars, the current data are consistent with the hypothesis that the flares originate in the stellar magnetospheres. 

\end{abstract}

\keywords{Early-type stars, Magnetic stars, Magnetospheric radio emissions, Radio transient sources, Transient detection}


\section{Introduction}\label{sec:intro}

Magnetic fields have been observed in planets, brown dwarfs, and stars of all spectral types, from ultracool dwarfs to the very hot massive stars (spectral types O, B and A, or early-type stars) \citep[e.g.][etc.]{grunhut2012,shultz2018,hallinan2006,kao2018, zarka2004}. The characteristics of stellar magnetic fields vary with spectral type, which is often attributed to the origin of the field in a given class of star. For a Sun-like star, the field is produced and sustained by a convective dynamo operating close to the surface of the star \citep[e.g.][]{tobias2002}. These fields are highly dynamic and have complex topologies. They give rise to explosive phenomena, such as flares and coronal mass ejections (CMEs). These phenomena are collectively referred to as stellar activity. Note that all stellar activity is inherently associated with some change in the magnetic configuration (reconnection of field lines).

By contrast, the magnetic OBA stars, which constitute $\gtrsim 7\%$ of the early-type star population\footnote{the recent study by \citet{hubrig2023} suggests magnetic fields may be universal in O-type multiple systems} \citep{grunhut2012,grunhut2017,sikora2019, hubrig2023}, are thought to have quieter surroundings. This is due to the fact that their magnetic fields are extremely stable throughout their lifetimes, and typically have simple topologies \citep[e.g.][]{kochukhovetal2019}. In most cases, the magnetic fields in OBA stars can be described as a dipole tilted to the rotation axis \citep[][etc.]{stibbs1950,shultz2018,sikora2019}. The interaction of these stable magnetic fields with the stellar winds results in various magnetospheric phenomena that exhibit only periodic variability due to rotational modulation arising from the misalignment between rotational and magnetic axes \citep[e.g.][]{oksala2015,leto2017}. This predictability represents a fundamental difference from the random stellar activity of late-type stars.

In recent years, ideas regarding the magnetospheric dynamics in rapidly rotating and strongly magnetic hot stars have undergone revolutionary changes. Such stars have magnetospheres in which the Kepler radius, $R_\mathrm{K}$ (the distance at which the centrifugal force due to co-rotation balances gravity), is smaller than the Alfv\'en radius, $R_\mathrm{A}$ (the radius of the largest closed magnetic field line). The region inside $R_\mathrm{K}$ is called the Dynamical Magnetosphere, and that between $R_\mathrm{K}$ and $R_\mathrm{A}$ is called the Centrifugal Magnetosphere \citep[DM and CM respectively,][]{petit2013}. 

In the DM, the radiatively driven stellar wind material follows the magnetic field lines, and is then drawn back to the star (again following the field lines) by gravity. Inside the CM however, the outward centrifugal force is stronger than gravity, and it is the magnetic tension that keeps material from escaping the stellar magnetosphere. By considering a rigidly rotating magnetosphere, \citet{townsend2005} showed that within the CM, there are certain locations along the field lines where plasma can stably accumulate. The scenario of stable plasma clouds was successfully used to explain the observed periodic variability of different magnetospheric phenomena \citep[e.g.][]{oksala2015}. The only limitation of this scenario was that it did not explain what balances the continuous accumulation of stellar wind plasma inside the magnetosphere, since the magnetic field has only a finite capacity to hold material against the centrifugal force. 

The first explanation proposed was that once that stage is reached, a large-scale explosion takes place in which the magnetosphere opens up and all the confined plasma escapes \citep{townsend2005}. This phenomenon was named centrifugal breakout (CBO) and predicted a highly variable magnetosphere surrounding these stars in which large-scale explosions occasionally occur. But no variability in the optical light curve other than that incurred due to rotational modulation was detected; nor were systematic changes in the depths of the light minima due to reductions in the magnetospheric column density \citep{townsend2013}. 

An alternate explanation invokes more complex physical processes such as diffusion of materials through the magnetosphere \citep{owocki2018}. This scenario predicts that the properties of different magnetospheric phenomena (such as the emission strength of the H$\mathrm{\alpha}$ emission) are dependent on the stellar mass-loss rate or its proxy. In 2020, by conducting comprehensive examination of $\mathrm{H\alpha}$ emission properties for stars with CM, \citeauthor{shultz2020} showed that this was not true.

The indifference of the magnetospheric phenomena to the mass-loss rate is, however, easily explained in the CBO scenario. This realization led to the new proposition that massive star magnetospheres are always maintained at the highest possible plasma density such that CBOs occur continuously, not as one single large-scale explosion that empties the magnetosphere but rather as a collection of small-scale eruptions occurring along different magnetic azimuths \citep{shultz2020,owocki2020}. Note that this study did not include late B-type or A-type stars as such stars are not usually found to be bright in H$\alpha$, presumably due to their weak stellar wind, or metallic wind \citep[][and references therein]{shultz2020}. This raises the possibility that the cooler early-type stars might differ from their hotter counterparts in terms of magnetospheric operation and phenomena, and CBO may not be relevant for late B/A-type stars. 

This was, however, found to be unlikely when \citet{leto2021} and \citet{shultz2022} showed that the incoherent radio emission from early-type stars, including down to A0, follows the same scaling law with stellar parameters. This observation can be satisfactorily explained by invoking CBO driven magnetic reconnections \citep{owocki2022}.

Thus, the most recent idea is that centrifugal magnetospheres of OBA stars are not quiet, but that they actually experience numerous tiny explosions leading to the ongoing escape of the magnetospheric plasma. Because we cannot resolve these objects, the signature of these continuous explosions, distributed randomly around the stellar magnetosphere, get washed out, giving the deceptive appearance of a stable magnetosphere. 

Observationally, evidence for these small-scale explosions has been limited.  Due to its high directivity and consequent sensitivity to changes on small spatial scales, the detection of electron cyclotron maser emission (ECME) is one possibility \citep{das2021}.  Alternately, it is tempting to extrapolate from solar flares, for which the occurrence frequency of the flares follows a power law with flare energies such that weaker flares are more frequent than stronger flares \citep[see Figure 10 of ][]{aschwanden2000}. If a similar case holds for CBOs in hot star magnetospheres, we would expect that there will be some instances when a strong explosion will occur that will not get washed out post spatial averaging, and hence will be detectable as a flare.

In the past, there have been reports of observation of flares from massive stars in various wave-bands such as optical, X-ray and radio (\S\ref{sec:flare_claims}). Some of these stars were subsequently discovered to have low-mass companions which might have hosted the flare \citep[e.g.][]{bouy2009,pedersen2017}. Even for the cases when no companion was detected, the flares were considered the manifestation of an otherwise invisible companion \citep[e.g.][]{naze2014}. The key reason behind this was that there was no theoretical explanation for the production of flares in hot star magnetospheres.

With the new understanding provided by the CBO mechanism it is now important to revisit the idea of whether hot magnetic stars flare or not. Since one cannot rule out the possibility of a companion star or a foreground object for the individual cases, the only way to understand whether or not hot magnetic stars flare is via statistical analysis. This paper describes the first attempt to achieve that goal. 

In this work, we present the first results from a search for radio flares from hot magnetic stars using the Very Large Array (VLA) Low-band Ionosphere and Transient Experiment (VLITE) Commensal Sky Survey (VCSS)\footnote{http://vlite.nrao.edu}. This paper is organized as follows. In Section~\ref{sec:flare_claims} we review previous claims of flare detections from hot (OBA) stars. Section~\ref{sec:strategy} describes our strategy for probing transient phenomena from hot magnetic stars. A description and analysis of VCSS data is given in Section~\ref{sec:vcss}. Section~\ref{sec:results} presents a discussion of our results with a summary given in Section~\ref{sec:summary}.

\section{Previous Claims of detection of flares from hot magnetic stars}\label{sec:flare_claims}

X-ray flares have been reported from both magnetic (field strength $\gtrsim 100$ G) and non-magnetic/weakly-magnetic hot stars.
One of the earliest observations of flares from the direction of a magnetic early-type star was made by the ROSAT survey in February 1995 for the magnetic B star $\sigma$ Ori E  \citep{groote2004}. From the same star, another X-ray flare was reported by \citet{pallavicini2002} and \citet{sanz-Forcada2004} using the $XMM-Newton$ telescope in March 2002. Subsequently however, a low-mass companion was discovered by \citet{bouy2009}, and the previously observed flares were attributed to the companion as the latter belongs to an established class of flare emitters. This was disputed by \citet{mullan2009}, who considered the low possibility of occurrence of two large flares on a late-type star within a span of just seven years. \citet{petit2012} observed the star at higher spatial resolution with the $Chandra$ X-ray Observatory, and found the low-mass companion to have significant contribution ($40\%$) to the overall X-ray flux observed from the system. This discovery proved that the low-mass companion produces significant amounts of X-rays and, as a result, made the claim of $\sigma$ Ori E as the origin of the flare more doubtful. 

In addition to $\sigma$ Ori E, there are two other hot stars, both with $\sim$kG strength surface magnetic fields (Shultz et al. in prep.) and not known to have low-mass companions, from which X-ray flares were reported. These two stars are IQ Aur \citep{robrade2011} and HD~47777 \citep{naze2014}. In both cases the presence of an undiscovered low-mass companion was speculated to be the source of the observed flares.

Apart from X-rays, the only other wave-band where a flare has been observed from the direction of a \textit{magnetic} massive star is radio. \citet{trigilio2000} first discovered that the magnetic early-type star CU\,Vir produces periodic radio pulses by ECME. In the same observations they also observed secondary enhancements at 1.4 GHz that could not be confirmed as persistent features. In 2021, \citeauthor{das2021} reported ultra-wideband radio observations of the same star that cover the frequency range of 0.4--4.0 GHz. In addition to the expected periodic radio pulses, they also observed strong (comparable to the persistent radio pulses), highly circularly polarized enhancements at rotational phases much offset from those of the persistent radio pulses at sub-GHz frequencies. Their subsequent observations conducted after a year confirmed these enhancements to be transients. This is the first (and only) confirmed observation of radio flares from the direction of a magnetic massive star.

\section{Strategy}\label{sec:strategy}

Our aim is to understand whether or not hot magnetic stars can exhibit transient phenomena (flares). There are two ways to attempt to achieve that goal: the first is to monitor a selected sample of hot magnetic stars (targeted observations), and the other is to look for such flares in existing databases from all-sky surveys. For our science, we deem the latter to be more efficient since it will include a much larger number of hot magnetic stars than what is possible through targeted observations.

There are two wavebands in which flares from hot magnetic stars have been claimed in the past: X-ray and radio. As a first step we chose to work with radio surveys. A number of large area radio surveys are either ongoing or recently completed over a wide range of frequencies, including the VLA Sky Survey \citep[VLASS, 3 GHz;][]{lacy2016}, ThunderKAT survey \citep[1.8 GHz,][]{fender2016}, all sky circular polarization survey with the Murchison Widefield array \citep[200 MHz,][]{lenc2018}, LOw Frequency ARray Two-meter Sky Survey \citep[LoTSS][]{shimwell2019}, TIFR Giant Metrewave Radio Telescope Sky Survey Alternative Data Release 1 \citep[TGSS ADR1, 150 MHz;][]{tgss}, Rapid ASKAP Continuum Survey \citep[RACS, 887.5 MHz;][]{mcconnell2020}, and the ASKAP Variables and Slow Transients \citep[VAST, $\gtrsim 1$ GHz,][]{murphy2013}.

While we do not have information to prefer one frequency range over another, we consider two important aspects that have been recently reported regarding non-thermal radio emission from hot magnetic stars. The first is that \citet{leto2021} discovered incoherent radio emission from hot magnetic stars have spectra that exhibit turn-over at $\sim 1$ GHz and the flux density decreases as one goes to frequencies below 1 GHz. This is consistent with the low detection rate of Ap--Bp stars in the LoTSS survey \citep{hajduk2022}. On the other hand, for coherent radio emission from hot magnetic stars, the sub-GHz frequency range above 200 MHz appears to be more preferred than the GHz regime \citep{das2020b,das2021,das2022b,das2022c}. In addition, the only confirmed observation of a radio flare from the direction of a hot magnetic star was obtained at 0.7 GHz \citep{das2021} and a tentative flare at 0.4 GHz was also reported by \citet{das2021}. Considering these aspects, we chose to work with a survey that operates above 200 MHz and below 1 GHz. The VCSS survey meets this criterion. We give details of VCSS in the next section.

We use a catalog of magnetic hot stars assembled from an exhaustive literature compilation (Shultz et al.\ in prep.). The catalog includes 761 stars of spectral types O, B, and A with at least one confirmed magnetic detection. Stars on the main sequence, pre-main sequence, and post-main sequence are included. The catalog also provides effective temperatures, luminosities and rotation periods, where available. This information is used to assess whether stars with observed transient radio emission have centrifugal magnetospheres necessary for the bursts to originate from CBO-driven flares.

Note that in the event of a flare detection, our current approach is not sufficient to discern whether that flare originated at the stellar magnetosphere or a magnectically active companion, but it does provide candidates that will be objects of interest for follow-up targeted observation campaigns.

\section{The VCSS data}\label{sec:vcss}

VLITE is a commensal instrument on the Karl G.\ Jansky VLA developed by the Naval Research Laboratory \citep[NRL;][]{clarke2018}. VLITE collects data at frequencies between 320--384~MHz from the low frequency prime focus receivers during nearly all regular VLA operations at GHz frequencies. VLITE data are recorded on up to 18 antennas, correlated, and transferred to NRL for automated pipeline processing and archiving. Removal of persistent radio frequency interference in the upper portion of the VLITE band gives an effective bandwidth $\sim 40$~MHz centered on 340~MHz.

A special correlator mode was enabled to allow VLITE processing of the on-the-fly data recorded during observations for VLASS. VLASS is a multi-epoch survey observing the $\delta > -40^{\circ}$ sky three times starting from September 2017. We use VLITE data from the first two observing epochs in this work. 

VLASS tessellates the sky with observing tiles covering $10^{\circ} \times 4^{\circ}$ in right ascension (RA) and declination, respectively. Tiles are observed by scanning in RA along lines of constant declination spaced by $7.2^{\prime}$, the half power radius of the VLA primary beam at 3~GHz. VLITE data are allowed to accumulate using a standard 2\,s sampling as the antennas move through an angular distance of $1.5^{\circ}$. This corresponds to $\sim28$\,s of time over most of the sky for the standard VLASS slew rates. These data are then correlated and imaged using the midpoint of the motion as the phase center. The result is short ``snapshots" with a smeared or elongated primary beam response. The \citet{pb2017} fluxes for 3C138 and 3C286 are used to set the flux scale. The calibrated and imaged VLITE data recorded during VLASS define VCSS.

Approximately $160,000$ snapshot images covering a $\sim 32,000$~deg$^2$ sky footprint are produced during each VCSS epoch. Snapshots are $3.5^{\circ}$ across with a typical sensitivity $7-10$~mJy~bm$^{-1}$ that increases away from image center due to primary beam attenuation. The attenuation is about a factor of three at the edge of the field of view (FoV).

VCSS snapshots are highly overlapping due to VLITE's wide FoV ($3.5^{\circ}$) relative to the declination spacing ($7.2^{\prime}$) of the VLASS observing strips. As a result, each point on the sky is typically sampled by $40-50$ snapshots spread over $80-100$~min, although irregular sampling over several months to a year occurs along tile boundaries. 

Snapshots in the first survey epoch are imaged with elliptical synthesized beams with axes ranging $7-35^{\prime\prime}$. In the second epoch, snapshots are imaged with $20^{\prime\prime}$ round beams to facilitate combination into more sensitive mosaic images. While these are useful to survey for long timescale transients, combining snapshots imaged over hours to months suppresses short timescale emission typical of stellar flares. For this reason, we do not use the mosaic images in this work and focus our search on the VCSS snapshots.

VCSS snapshots are processed with the VLITE Database Pipeline \citep[VDP,][]{vdp}. The Python Blob Detector and Source Finder \citep[PyBDSF,][]{pybdsf} is used to catalog sources with signal-to-noise ratios (SNR) greater than five, where the noise is calculated locally around the source position. 

Querying the VDP database for snapshot sources within $10^{\prime\prime}$ (the approximate minimum VCSS resolution) of hot magnetic stars from the catalog revealed three matches. Each VCSS match appears in only one snapshot and none correspond to known radio sources. However, the VCSS survey is known to contain a large number of imaging artifacts. More than one million sources in each epoch are detected only once and unassociated with known radio sources. We therefore performed an analysis to determine the likelihood our star detections are due to false associations.

\subsection{VCSS Artifacts}

In Appendix~\ref{app:art}, we provide an in-depth analysis of artifacts present in VCSS snapshots. Our findings reveal the existence of two distinct artifact populations: one concentrated within an arcminute of bright sources, and another distributed uniformly in all directions. Given that the nearest neighbors for the three star detections are situated more than 500$^{\prime\prime}$ away, we concentrate our analysis on the isotropic artifact population.

We also observe a correlation between the density of artifacts and the SNR; fewer artifacts are observed at higher SNR. Additionally, the substantial variation in the total artifact count per snapshot indicates the number of false associations within any stellar catalog will depend on the detection threshold, the distribution of stars and the specific set of snapshots they encompass.

\subsection{False Detection Probability}

The probability, $P$, of observing $N$ events when $\lambda$ are expected is given by Poisson statistics:
\begin{equation}
   P(N;\lambda) = \frac{\lambda^N e^{-\lambda}}{N!} 
\end{equation}

The expected number of false associations in a single snapshot is determined by the product of the solid angle searched and the density of artifacts in the snapshot. All three VCSS detected hot magnetic stars have SNR~$> 5.5$. Therefore, we calculate the product of the $10^{\prime\prime}$ search radius solid angle, $\Omega_*$, with the isotropic artifact density above a SNR of 5.5 for each snapshot, $n/\Omega_{FoV}$, where $\Omega_{FoV}$ is the snapshot solid angle (9.6~deg$^{2}$). We sum over all stars and all snapshots they are situated in to get the expected number of false associations for the catalog:
\begin{equation}
    \lambda = \sum_{j}^{stars} \; \sum_{i}^{snaps} \Omega_* \; \frac{n_i}{\Omega_{FoV}}
\end{equation}
We calculate $\lambda=0.44$ for our star catalog. The probability of three false associations when 0.44 are expected is $0.9\%$. We confirm this result with simulations of randomly generated star catalogs in Appendix \ref{app:sim}.

\section{Results and Discussions}\label{sec:results}

\begin{figure}[htb]
    \centering
    \includegraphics[width=3in]{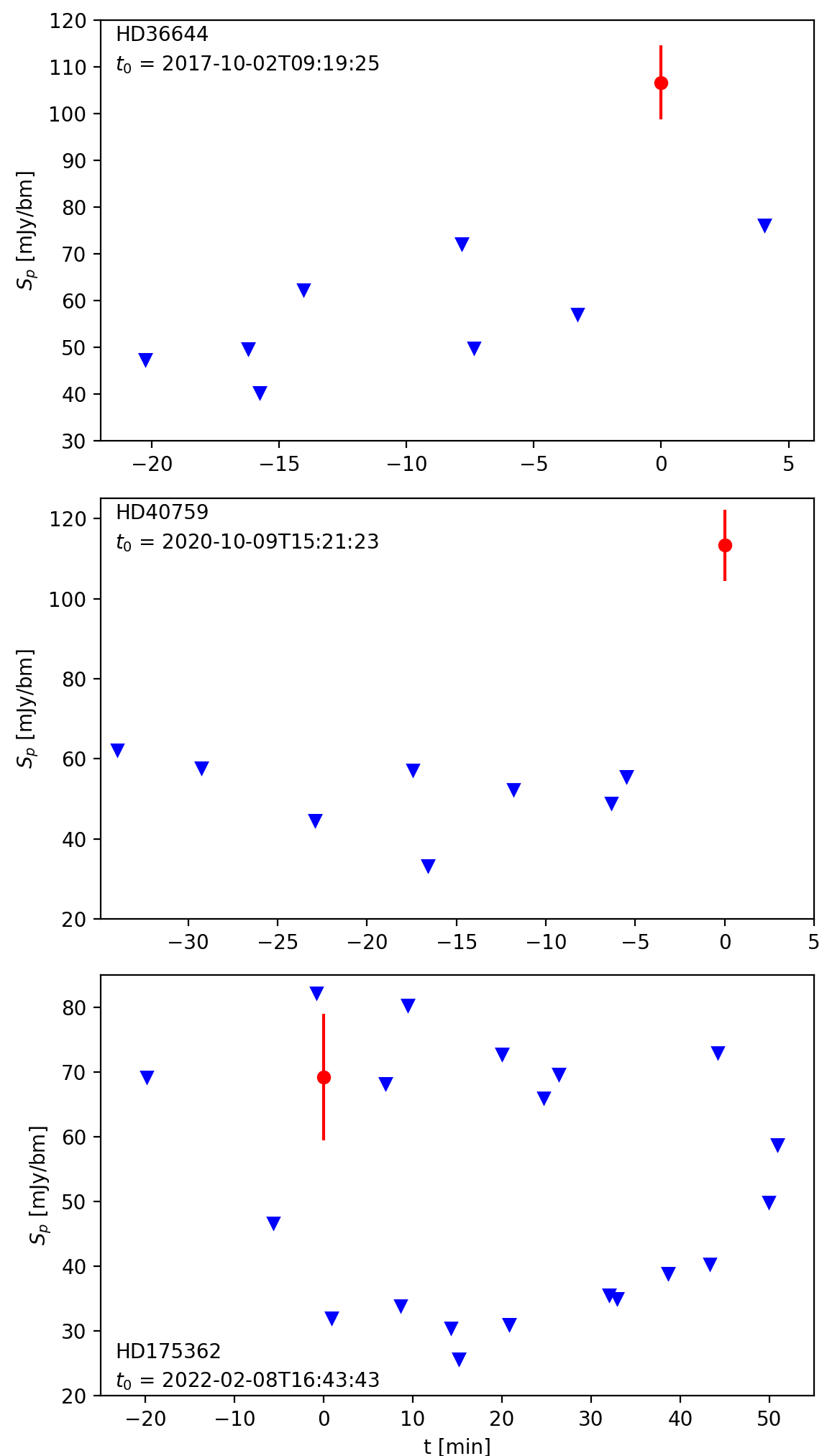}
    \caption{Light curves of hot magnetic stars detected in VCSS snapshots. The time axis is aligned with the UT of each detection, indicated and labeled in each plot. Upper limits for non-detections (triangles) correspond to three times the PyBDSF-calculated local noise at the stellar positions corrected for primary beam attenuation}. \label{fig:lcs}
\end{figure}

The three stars detected by VCSS are HD\,36644, HD\,40759 and HD\,175362. Figure~\ref{fig:lcs} presents the corresponding light curves for these stars. Brightness errors include $1\sigma$ fitting uncertainties with $3\%$ primary beam correction uncertainty added in quadrature but do not include the $15\%$ VLITE flux scale calibration uncertainty.

Upper limits from non-detections are defined as three times the local noise. We use the local noise maps calculated by PyBDSF across the FoV corrected for primary beam attenuation. Implementing rigorous quality controls, VDP prioritizes the flagging of poor quality images, even if it results in a small fraction of quality images being flagged. In cases where our examination indicated reliable flux measurements from flagged images, their upper limits have been included in Figure~\ref{fig:lcs}. A discussion of each detection follows.

HD~36644 was detected on 2017 October 2 at 09:19:25 UT during a 28\,s duration snapshot. The brightness was $107 \pm 8$\,mJy\,bm$^{-1}$ with a SNR of 5.6. Its nearest neighboring source was 918$^{\prime\prime}$ away. Non-detections effectively constrain the emission timescale to under seven minutes. 

HD~36644 lay beyond the northern border of the tile under VLASS observation on the day of the detection. The stellar position, however, falls within the coverage of nine VCSS snapshots spanning six VLASS declination strips. These strips were observed sequentially from north to south, resulting in a general trend of HD~36644 appearing farther out in the VCSS primary beam as time progressed. The larger beam attenuation at larger distances creates the upward drift in the non-detection limits in Figure~\ref{fig:lcs}.

\begin{figure*}[ht]
    \centering
    \includegraphics[width=6.5in]{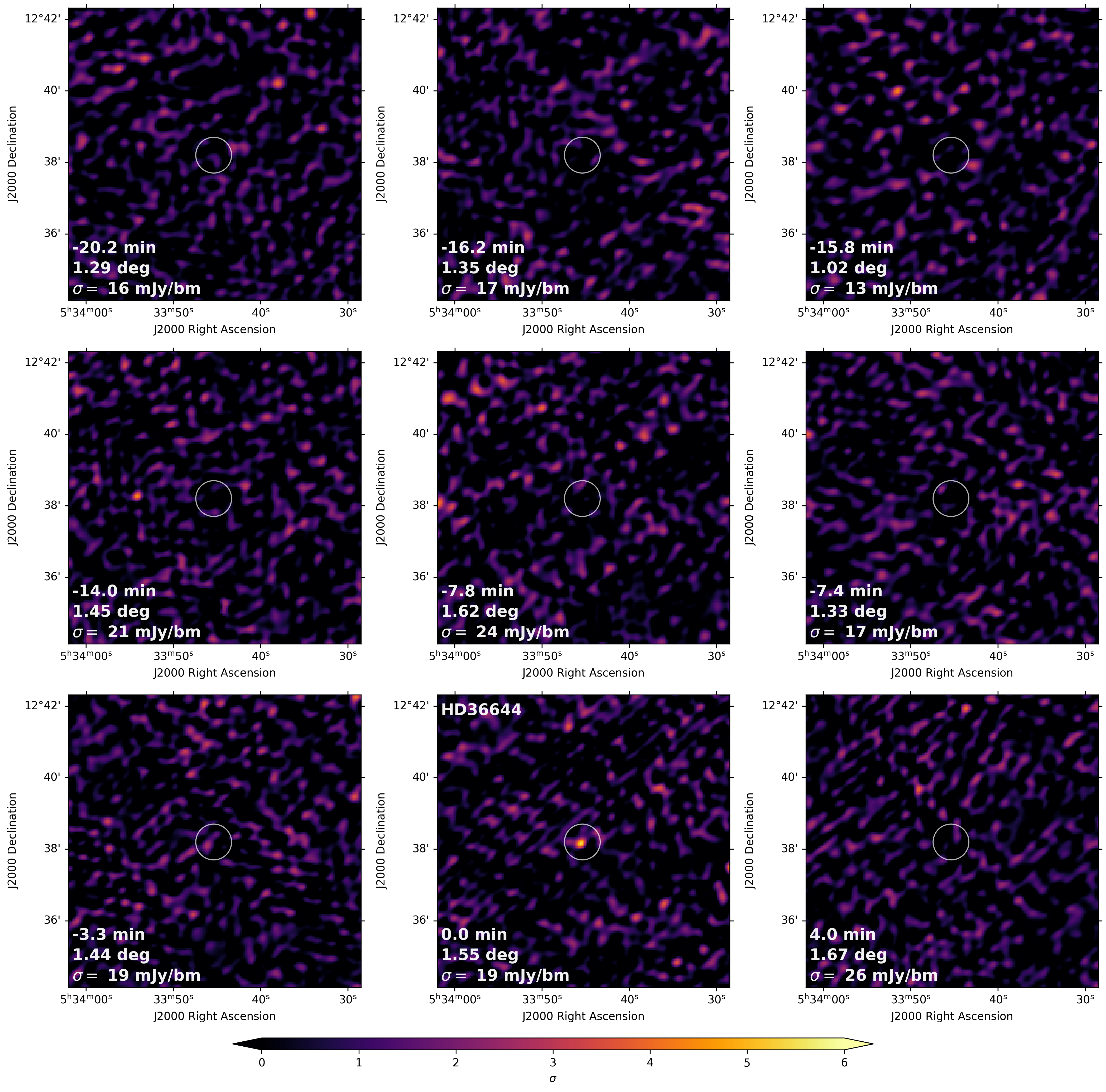}
    \caption{VCSS snapshot cutouts centered on HD~36644, organized chronologically. Pixel intensities have been corrected for primary beam attenuation. Labels indicate the time relative to the detection, the distance of HD~36644 from the snapshot center and the noise ($\sigma$) at the stellar position (circled). The color scale ranges 0-6 times the local noise for each image.}
    \label{fig:stamps}
\end{figure*}

For illustrative purposes, Figure~\ref{fig:stamps} presents cutout images centered on the position of HD~36644 at the times indicated in the light curve. Notably, intensity noise at the stellar location increases as the distance from the snapshot center grows, primarily due to the effects of primary beam attenuation.

\begin{figure}[htb]
    \centering
    \includegraphics[width=3in]{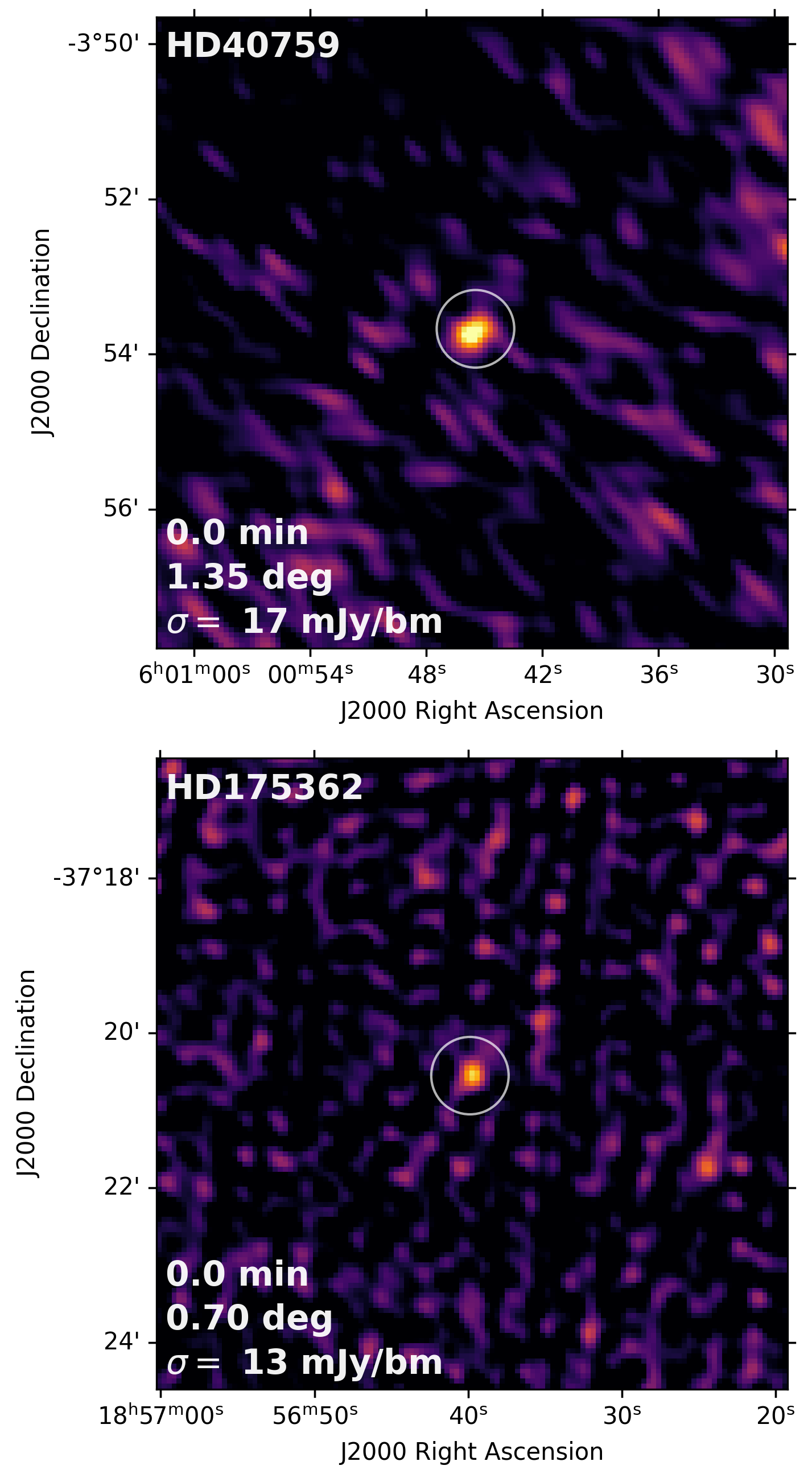}
    \caption{VCSS snapshot cutouts centered on HD~40759 ({\it top}) and HD~175362 ({\it bottom}) at the time of detection. Color scale is as in Figure~\ref{fig:stamps}. \label{fig:stamps2}}
    \label{lcs}
\end{figure}

HD~40759 was detected with a significance of $6.8\sigma$ on 2020 October 9 at 15:21:23 UT. Similar to the previous case, this detection occurred in a 28\,s snapshot, with the source positioned 1248$^{\prime\prime}$ from its nearest neighbor. Its observed brightness was $113 \pm 9$\,mJy\,bm$^{-1}$. The upper panel of Figure~\ref{fig:stamps2} presents the image cutout captured at the time of detection.

During the time of its detection, the star was situated off the northwest corner of the tile under observation by VLASS. Over time, observations along the strips proceeded south to north, gradually approaching the stellar location. The decreasing primary beam attenuation creates the downward trend in non-detection limits within the HD~40759 light curve. While the duration of the emission remains unconstrained, the non-detections preceding the burst effectively impose an upper limit on the rise time of $\sim 5$~minutes.

Unlike the other two stars, HD~175362 was detected within the tile being observed by VLASS. The temporal pattern of non-detections with values below $\sim 50$~mJy~bm$^{-1}$ exhibits a distinctive sequence of decreasing, flattening, and then increasing. This pattern corresponds to the primary beam attenuation sampled across the north-south direction of the VLASS declination strips by snapshots aligned with the stellar RA. Additionally, the larger non-detection values occurring at all times reflect the larger primary beam attenuation of snapshots centered at large distances in the east-west direction along the observing strips.

The detection of HD~175362 occurred on 2022 February 8 at 16:43:43 UT during a 42\,s snapshot. The fitted brightness was $69 \pm 10$\,mJy\,bm$^{-1}$, 5.5 times the local noise level. Positioned 529$^{\prime\prime}$ away from its nearest neighbor, this detection is flanked by non-detections before and after it, effectively constraining the emission timescale to less than $\sim$ 1 minute. To further illustrate this event, the lower panel of Figure~\ref{fig:stamps2} presents the image cutout at the moment of detection.

Although none of the stars are detected by VLASS, all VCSS detections occurred beyond the narrow VLASS observing strip coincident with each snapshot. Specifically, HD~36644 remained unobserved by VLASS until 10 days after its VCSS detection. Similarly, HD~40759 had been observed by VLASS 82 days prior to its VCSS detection. The VCSS detection of HD~175362 preceded the VLASS observation of the star's position by $\sim 20$ minutes. However, the VCSS upper limit on the burst timescale is considerably shorter than the VLASS observing delay. 

In the following subsections, we discuss the plausibility of these stars serving as potential sources of the observed radio transients.

\subsection{Do the stars have centrifugal magnetospheres?}\label{subsec:centrifugal_or_not}

As described in the introduction, CBOs are the only known phenomena that could give rise to an observable transient event from a hot magnetic star without any influence of a companion. But for this scenario to be valid, the stars must have centrifugal magnetospheres (CMs). In this subsection, we examine whether the three stars harbor CMs. 

The two parameters needed to examine this property are the Kepler radius $R_\mathrm{K}$ and the Alfv\'en radius $R_\mathrm{A}$.
$R_\mathrm{K}$ can be calculated using the following equation;
\begin{align}
    R_\mathrm{K}&={\left(\frac{GM_*}{\Omega^2}\right)}^\frac{1}{3}\label{eq:kepler_radius}
\end{align}
where, $G$ is the universal gravitational constant, $M_*$ is the stellar mass and $\Omega=2\pi/P_\mathrm{rot}$ is the stellar rotational frequency.

$R_\mathrm{A}$, on the other hand, is a more complicated function of several stellar parameters such as the stellar magnetic field strength, rotation period, mass-loss rate, wind speed and also the misalignment between rotation and magnetic axes (obliquity $\beta$). $R_\mathrm{A}$ increases with increasing magnetic field, rotation period, and decreases with increasing mass-loss rate. Barring the case of the very rapid rotators ($P_\mathrm{rot}\lesssim$ a day), $R_\mathrm{A}$ decreases with increasing stellar wind-speed, and is nearly independent of $\beta$. For the very rapid rotators, $R_\mathrm{A}$ can increase with increasing stellar wind-speed and increasing $\beta$ \citep[e.g. see \S 2.2 of ][]{trigilio2004}. As can be seen below, none of the three stars under consideration are very rapid rotators, so that we can ignore the dependence of $R_\mathrm{A}$ on $\beta$ and the rotation period. In that case, the equation for $R_\mathrm{A}$ can be approximated by the following \citep[for dipolar magnetic fields,][]{ud-doula2008}:
\begin{align}
    \frac{R_\mathrm{A}}{R_*}&=0.3+{(0.25+\eta_*)}^\frac{1}{4} \label{eq:alfven_radius}\\
    \eta_*&=\frac{B_\mathrm{eq}^2R_*^2}{\dot{M}v_\infty}\nonumber
\end{align}
where $R_*$ is the stellar radius, and $\eta_*$ is the `magnetic confinement parameter' \citep{ud-doula2008} that is a function of the surface equatorial magnetic field $B_\mathrm{eq}$, $R_*$, mass-loss rate $\dot{M}$ and wind terminal speed $v_\infty$.

Among the three stars, HD\,175362 is the most well-studied. It is also the hottest star among the three \citep[$T_\mathrm{eff}=17.6\pm0.4$ kK,][]{shultz2019b}. The star has a magnetic field with polar strength of $\approx 17$ kG \citep{shultz2019c}. Its rotation period was estimated as  3.67381(1) days \citep{bohlender1987,shultz2018}, where the uncertainty in the least significant digit is in brackets. The stellar radius and mass were estimated to be $R_*=2.7\pm0.2\,R_\odot$ and $5.3\pm0.2\,M_\odot$ \citep{shultz2019c}, respectively, which gives a Kepler radius of $6.5\pm0.4\,R_*$. This is much smaller than the reported Alfv\'en radius of the star which is $\approx 46\,R_*$ \citep{shultz2019c}. Thus the star harbors a centrifugal magnetosphere, and hence may experience CBO phenomena. This is also supported by the fact that the star's non-thermal gyrosynchrotron radio luminosity follows the scaling relation with stellar parameters expected for CBO driven radio emission \citep{shultz2022,owocki2022}. 

The second best-studied star is HD\,40759, a de-facto spectroscopic triple-star system with all three components being A1--B9 stars. The effective temperature $T_\mathrm{eff}$ of the components lies in the range 9--12 kK (Semenko et al. in prep). The magnetic component is a chemically peculiar star with a rotation period of 3.37484 days \citep{semenko2022}. The maximum observed longitudinal magnetic field is $\approx 2$ kG, with a root mean square value of 1.12 kG \citep{romanyuk2021, semenko2022}. Accurate mass and radius estimates for the star are not available at the moment, yet we can evaluate them roughly from the evolutionary status of the star. HD\,40759 is a member of the stellar association Orion OB1, subgroup 1c, with an average age of 4.6~Myr \citep{semenko2022}. Interpolating stellar evolutionary tracks provided by the MIST project \citep{mist0, mist1}, gives the mass and radius of the star as 2.5\,$M_\odot$ and 2.1\,$R_\odot$. Thus, we obtain $R_\mathrm{K} =  6.1 R_*$.

From the lower limit of the surface magnetic field (2 kG), we can also estimate the Alfv\'en radius by assuming typical values of mass-loss rate ($\dot{M}$) and wind terminal speed ($v_\infty$) observed for late B/A-type stars. We take $\dot{M}=10^{-10}\,M_\odot/\mathrm{yr}$, $v_\infty =1000$ km/s, which gives {\bf $R_\mathrm{A}\approx 14\,R_*$}. Note that this value is very likely a lower limit to the true Alfv\'en radius since the value used for the surface magnetic field is actually a lower limit and that used for $v_\infty$ is likely an upper limit to the respective true values ($R_\mathrm{A}$ increases with increasing surface magnetic field strength, and decreases with increasing $v_\infty$, see Eq. \ref{eq:alfven_radius}). Thus $R_\mathrm{A}>R_\mathrm{K}$ for this star as well, and, like HD\,175362, it possesses a centrifugal magnetosphere.  

\begin{figure*}[htb]
    \centering
    \includegraphics[width=\textwidth]{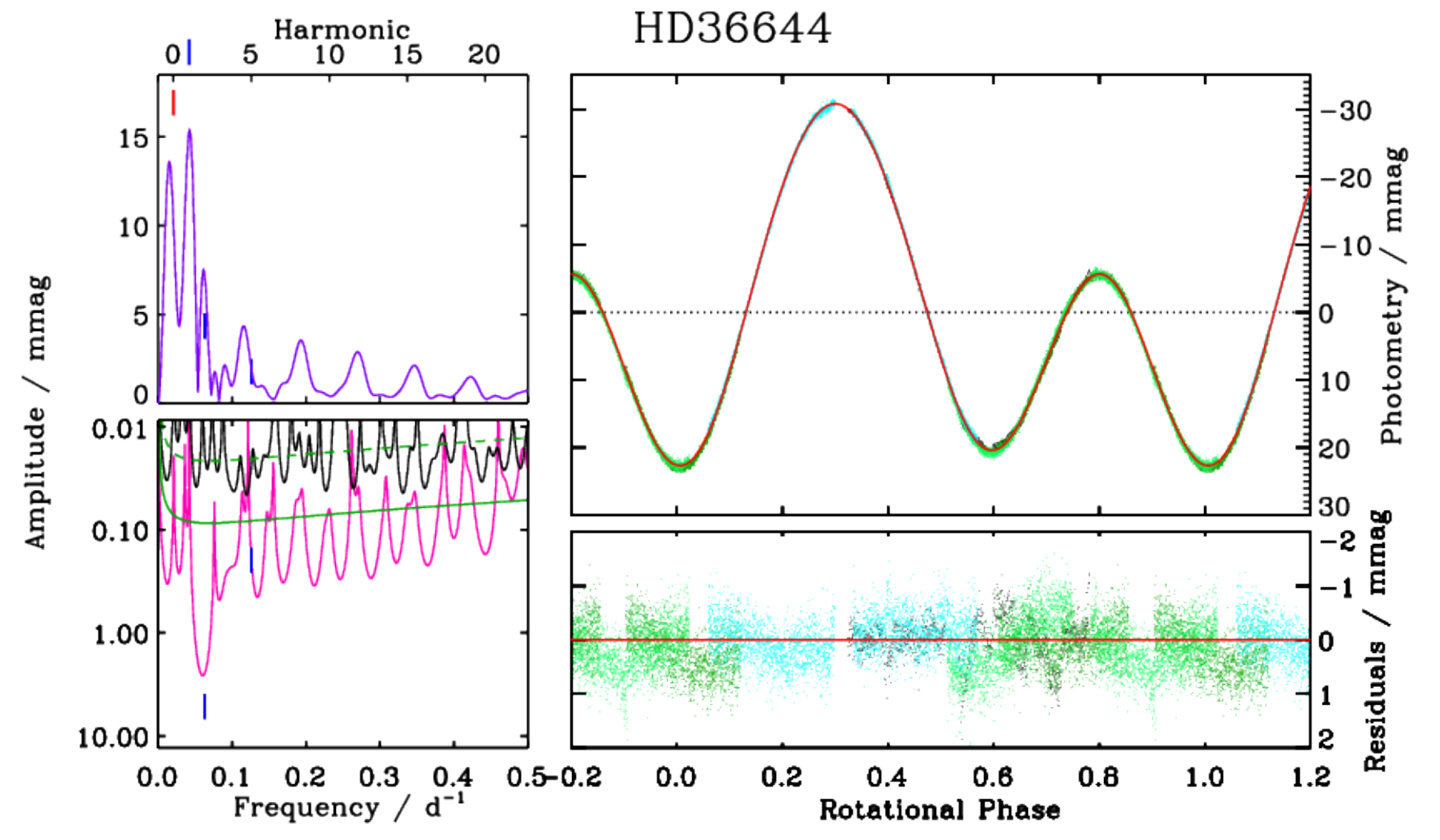}
    \caption{{\em TESS} photometry of HD\,36644. {\em Left panels}: periodograms for the full dataset (top, purple), after pre-whitening with the rotational frequency and the first harmonic (bottom, magenta), and after pre-whitening with all harmonics (bottom, black). The rotational frequency is indicated with a red dash, harmonics with blue dashes. The green dot-dashed curve is a polynomial fit to the fully pre-whitened frequency spectrum, used to evaluate the noise floor; the green solid curve shows 4$\times$ the noise floor, the threshold for frequency significance. {\em Right panels}: the light curve folded with the rotational frequency (top), and residuals (bottom) after subtraction of the harmonic model determined from frequency analysis (red curve). Different sectors are indicated by colour.} \label{HD36644_periods_lc}
\end{figure*}

Finally, the star HD\,36644 has the least known stellar parameters. \citet{chojnowski2019} reported the star to have a very strong magnetic field with a magnetic field modulus of 18 kG, which serves as the lower limit to the actual maximum surface magnetic field strength. Shultz et al. (in prep.) estimated the star's bolometric luminosity and effective temperature to be $\log (L/L_\odot)=1.3$ and $\log (T_\mathrm{eff}/\mathrm{K})=3.95$ respectively, which gives a stellar radius of $1.9\,R_\odot$. 

To estimate the rotational period, we acquired the High-Level Science Product light curves from the {\em Transiting Exoplanet Survey Satellite} ({\em TESS}) via the Mikulski Archive for Space Telescopes (MAST). HD\,36644 was observed during sectors 6 (with 30-minute cadence) and 43, 44, and 45 (with 10-minute cadence). The {\em TESS} light curve is shown in Fig.\ \ref{HD36644_periods_lc}.

Inspection of the individual sectors revealed long-term variability, suggesting a rotation period longer than the $\sim 1$~month duration of each sector. Determining the period therefore requires combination of the data from multiple sectors. Since the flux in any given sector is normalized by the mean flux in that sector, this leads to systematic offsets in the flux levels. 

An iterative approach was taken to determine the period. First, the approximate period was determined using the frequency analysis package {\sc period04} \citep{2005CoAst.146...53L}. The light curves were then folded with this period, and individually adjusted by adding an arbitrary flux offset in order to bring them into alignment. The frequency analysis was then repeated in order to obtain a refined period and a harmonic model for the variation by inclusion of all significant frequencies in the periodogram (where `significant' is defined as 4$\times$ the noise floor). The harmonic model was then used to perform an empirical detrending, by subtracting the harmonic model from each sector, fitting a low-order polynomial to the residuals, and then subtracting the residuals from the original light curve. These de-trended light curves were then used to obtain the final period. 

The results are shown in Fig.\ \ref{HD36644_periods_lc}. The light curve displays a double-wave variation, with the strongest peak in the frequency spectrum at $2f_{\rm rot}$, together with two additional significant harmonics of $f_{\rm rot}$. The rotational frequency is $f_{\rm rot} = 0.0210638(3)$~d$^{-1}$, corresponding to a rotational period of $P_{\rm rot} = 47.4746(6)$~d.

If this rotational period is correct, the corresponding Kepler radius is 36\,$R_*$ assuming a mass of $2\,M_\odot$.  If we assume the other parameters to have the same value as those assumed for HD\,40759, the extremely strong surface field implies that the Alfv\'en radius is at least 39~$R_*$. Thus, despite the very long rotation period, it is likely that the star has a thin centrifugal magnetosphere. We emphasize that the lower limit for the Alfv\'en radius is likely very conservative, first because 18 kG is the lower limit for the surface magnetic field strength, and second because the mass-loss rate of $10^{-10} \dot{M}~{\rm yr}^{-1}$ is probably considerably higher than the actual mass-loss rate of this Ap star, which may easily be orders of magnitude lower. In the future, it will be important to obtain more accurate parameters for the star so as to confirm the presence (or absence) of a centrifugal magnetosphere surrounding the star.

\subsection{Flare energies}

The flux density of HD\,175362 at 50 cm is $\approx 0.54$ mJy \citep{shultz2022}, corresponding to an isotropic spectral luminosity (defined as $4\pi d^2 S$, where $d$ is the distance to the star and $S$ is the observed flux density) of $\sim 10^{16}$ ergs/s/Hz \citep[using the distance to the star as 153 pc,][]{gaia2018}. The isotropic spectral luminosity corresponding to the VCSS flux density is $\sim 2\times 10^{18}$ ergs/s/Hz, thus higher by a factor of 100 than the incoherent spectral radio luminosity. 

In the case of HD\,40759, the flux density of the flare observed from its direction is around 100 mJy, this corresponds to an isotropic spectral luminosity of $\approx 2\times 10^{19}$ ergs/s/Hz \citep[assuming stellar distance of 430 pc,][]{gaia2018}. Unfortunately, the star's basal radio luminosity is unknown so far, but the inferred spectral luminosity is much higher than that obtained for incoherent radio emission from any hot magnetic star \citep[e.g.][]{leto2021}. From the VLASS Quick Look image local noise map, the $3\sigma$ upper limit to the star's flux density at 3 GHz is 0.6 mJy.

For HD\,36644 also, the basal radio flux density is unknown. We obtained a VLASS $3\sigma$ upper limit of 0.4 mJy for its 3 GHz flux density. Using the stellar distance as 384 pc \citep{gaia2018}, the isotropic spectral luminosity corresponding to the flare detection ($\sim 100$ mJy) is $\approx 2\times 10^{19}$ ergs/s/Hz, which is again too high to be of incoherent origin.

\subsection{Rotational phases corresponding to the flare detection}
The only known mechanism for the production of coherent radio emission by hot magnetic stars is ECME, which gives rise to radio pulses at rotational phases around the magnetic nulls \citep[e.g.][]{trigilio2000}. In this subsection, we examine the stellar rotational phases corresponding to their radio detections.

\begin{figure}[htb]
    \centering
    \includegraphics[width=3in]{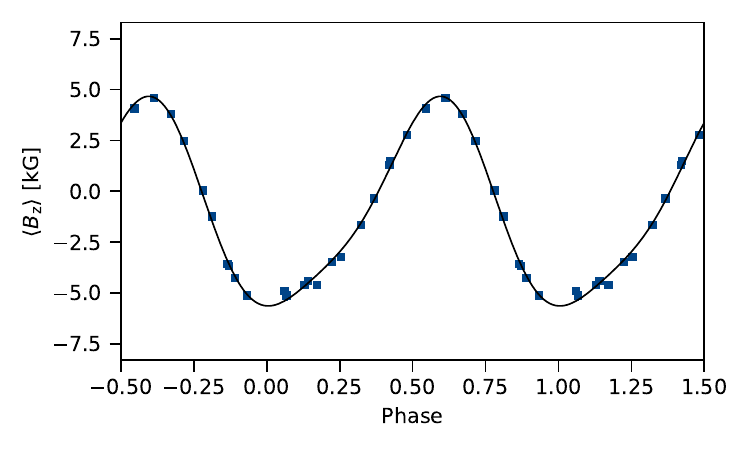}
    \includegraphics[width=3in]{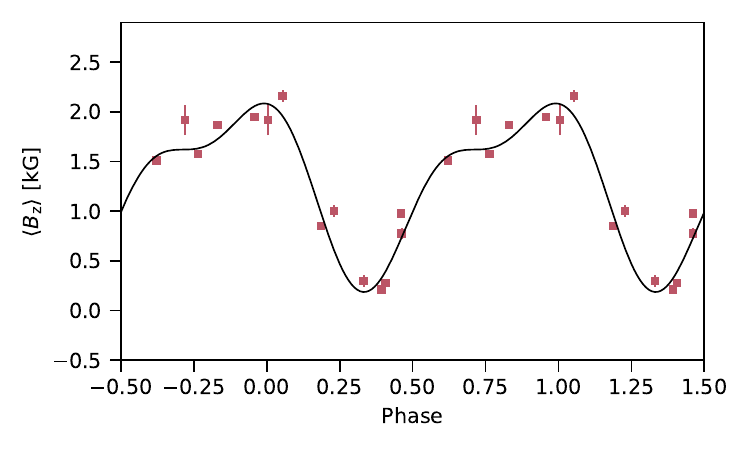}
    \caption{The longitudinal magnetic field $\langle B_\mathrm{z} \rangle$ curves for HD\,175362 (hydrogen lines, \citealt{shultz2018}, top) and HD\,40759 (metals, Semenko et al. in prep., bottom).
    \label{fig:bzcurves}}
\end{figure}

The HJD corresponding to the VCSS detection of HD\,175362 is 2459619.19268. Using the ephemeris of \citet{shultz2018}, we obtain the corresponding rotational phase as $0.16 \pm 0.02$. This rotational phase is significantly offset from the magnetic nulls at phases $0.38 \pm 0.01$ and $0.77 \pm 0.01$, but close to a \bz~extremum at phase 0 (Fig.\,\ref{fig:bzcurves}, top). In the case of HD\,40759, the \bz~curve never passes through zero though it approaches it very closely (Semenko et al. in prep.). Using the ephemeris: HJD = 2458487.0173 + 3.37484\,E (Semenko et al. in prep.), we obtain the rotational phase corresponding to our radio detection as 0.84. Interestingly, phase 0.84 is again close to a magnetic field extremum (Fig.\,\ref{fig:bzcurves}, bottom).

The VCSS detection of HD\,36644 occurred on HJD = 2458028.8900. Using the ephemeris from our {\em TESS} analysis in Sect.\,\ref{subsec:centrifugal_or_not}: HJD = 2459544.8(1) + 47.4746(6)\,E, we obtain the rotational phase as $0.07 \pm 0.01$. Unfortunately, the correspondence to the longitudinal magnetic field cannot be determined due to the current paucity of magnetic data for this star.

\subsection{Can the flares originate at the stellar magnetospheres?}

Under the scenario that the VCSS detections corresponds to flares from the stars, the underlying radio emission is almost certainly coherent in nature. Note that for CU\,Vir, in addition to the transient enhancements observed close to the extrema of the longitudinal magnetic field \bz, \citet{das2021} also observed a flux density enhancement of a magnitude (relative to the basal flux density of the star) similar to that observed for the stars reported here. This enhancement was observed at a rotational phase where regular pulses produced by ECME were expected. Hence, this was attributed to ECME, and the unusual strength of the enhancement was speculated to be related to an unusually strong CBO event \citep{das2021}. 

For the case of a star with an axi-symmetric dipolar magnetic field, ECME pulses ideally appear around the rotational phase corresponding to magnetic nulls. For the two stars for which we could correlate the rotational phases corresponding to the VCSS detections, the rotational phases are greatly offset from the phases of minimum $|\langle B_\mathrm{z}\rangle|$, and they rather lie close to the maxima of $|\langle B_\mathrm{z}\rangle|$. This situation is similar to the case of the radio flares observed from CU\,Vir \citep{das2021}, and supports the idea that they all represent a class of coherent transient events different from the periodic ECME phenomenon.

For the adopted distances, the estimated radio luminosities ($\sim 10^{18}-10^{19}$ ergs/s/Hz at 340 MHz) are some of the highest ever observed for stellar flares, and rules out certain alternative sources. M-dwarfs are one of the most common sources of stellar flares. From Figure 1 of \citet{vedantham2022}, M-dwarf flares have luminosities below $10^{15}$ ergs/s/Hz at 144 MHz. Thus M-dwarf companions to the stars are unlikely to produce flares such as the ones detected by VCSS. 

The only sources that produce flares above a luminosity of $10^{17}$ ergs/s/Hz (at 144 MHz) are the RS CVn systems and FK Com stars. So far, none of the stars are known to have any such companions. However, \citet{bodensteiner2018} reported the detection of an infrared nebula around HD\,175362, which they attributed to binary interactions. Thus, it is important to perform detailed follow-up observations of all three stars in the future in order to characterize their radio properties, and also at other wavebands to search for potential companions. 

In addition to being possible sources of the radio flares themselves, companions can also play an indirect role in the generation of flares from hot star magnetospheres in a manner similar to how Io triggers auroral radio emission from Jupiter \citep[e.g.][]{belcher1987}. In this case, when the companion moves through the magnetosphere of the hot star, it may lead to enhancement in the non-thermal electron density and manifest as a radio flare.

While flares of CBO origin will be random in terms of their occurrence times, those triggered by a companion are expected to exhibit a periodicity related to the orbital motion. With a single detection, however, it is not possible to distinguish between different scenarios. Follow-up observations must be performed, both for understanding the temporal properties as well as to detect companion(s). One of the key challenges for the latter will be to resolve objects that are only a few milli-arcseconds apart. In that case, one can perform Very Long Baseline Interferometry (VLBI) with telescopes such as the High Sensitivity Array (HSA) that offer angular resolution of $\lesssim 1$ mas. Even if the companion is not sufficiently radio bright during non-flaring phases, its presence might be detectable through positional shift of the hot magnetic star due to its orbital motion.

\section{Summary}\label{sec:summary}

In this work, we report radio flares at 340 MHz from the directions of three hot magnetic stars using VCSS data. By characterizing the imaging artifacts observed in VCSS snapshots, we find the probability that these three detections are false associations with artifacts is less than 1\%. This low probability motivates us to examine whether the stars can produce radio flares in their magnetospheres. Using the available stellar parameters, we conclude that all three stars are suitable for harboring centrifugal magnetospheres where CBOs can occur. Note that HD\,175362 is confirmed to have a centrifugal magnetosphere and HD\,40759 is extremely likely to have one. HD\,36644 needs to be investigated further to infer if it indeed possesses a centrifugal magnetosphere. We also find that similar to CU\,Vir (the only other hot magnetic star from which radio flares have been reported) the flares were observed close to a magnetic extremum, for the two stars with sufficient magnetic data. Whether this has implications for the underlying flare production mechanism can only be confirmed by more such detections.

To summarize, although our current data do not allow us to confirm the source of the observed enhancements in the radio light curves, neither does the data suggest the stars cannot be the origin of the radio flares. It will be interesting to see if more flares are detected from hot magnetic stars in the ongoing VCSS epoch 3, or by other radio surveys, since statistical analysis is probably the only way to confirm whether hot magnetic stars flare or not.

\begin{acknowledgements}

Basic research in radio astronomy at the U.S.\ Naval Research Laboratory is supported by 6.1 Base funding. Construction and installation of VLITE was supported by the NRL Sustainment Restoration and Maintenance fund. The National Radio Astronomy Observatory is a facility of the National Science Foundation operated under cooperative agreement by Associated Universities, Inc. BD acknowledges support from the Bartol Research Institute. This paper includes data collected with the {\it TESS} mission, obtained from the MAST data archive at the Space Telescope Science Institute (STScI). Funding for the TESS mission is provided by the NASA Explorer Program. STScI is operated by the Association of Universities for Research in Astronomy, Inc., under NASA contract NAS 5–26555. Support for MAST for non-HST data is provided by the NASA Office of Space Science via grant NNX13AC07G and by other grants and contracts. All the {\it TESS} data used in this paper can be found in MAST: \dataset[10.17909/t9-wpz1-8s54]{http://dx.doi.org/10.17909/t9-wpz1-8s54}

\end{acknowledgements}


\appendix

\section{VCSS Artifact Analysis} \label{app:art}

To investigate VCSS artifacts we first define a sample of isolated point sources (IPS) selected from large sky surveys with similar resolution to VCSS: the 150~MHz TGSS ADR1 \citep{tgss} and 887.5~MHz RACS \citep{racslo} catalogs.

We follow \citet{degas2018} in defining a source compactness metric, $C$, based on source SNR and total to peak flux ratio, $R = S_t/S_p$, to determine if sources are resolved or unresolved. $C = R_{2\sigma}/R$, where $R_{2\sigma}$ defines the flux ratio envelope below which $97\%$ of unresolved sources lie. Methods for defining this envelope are given by, e.g.~\citet{wenss}, \citet{bondi08} and \citet{tgss}. 

We adopt the form of the equation given by \citet{williams13} and performed our own fit to the TGSS ADR1 catalog:
\begin{equation}
    R_{2\sigma} = 1.071 + 2\sqrt{0.041^2 + (0.784 \; SNR^{-0.925})^2}    
\end{equation}
which agrees closely with the fit obtained by \citet{degas2018}. For RACS we use equation~(1) from \citet{racslo}:
\begin{equation}
    R_{2\sigma} = 1.025 + 0.69 \; SNR^{-0.62}
\end{equation}

Sources with $C \geq 1$ are point-like while $C < 1$ sources are assumed to be resolved. We defined IPS by selecting sources with $C \geq 1$ in both catalogs and having no neighbor within $300^{\prime\prime}$. IPS appearing in both catalogs were defined by cross-matching and selecting matched sources with separation $< 10^{\prime\prime}$. Our final sample includes 23,841 IPS spread across $8500$~deg$^{2}$ in the overlap footprint of TGSS, RACS and $\delta > -40^\circ$. 

We find 271,074 VCSS sources cataloged within $300^{\prime\prime}$ of 17,436 distinct IPS in 79,036 epoch 1 snapshots. In epoch 2, we find 230,255 sources around 15,900 distinct IPS in 74,912 snapshots. About $90\%$ of the sources in each epoch are VCSS detections of the IPS while about $10\%$ appear to be imaging artifacts.

\begin{figure}[ht]
    \centering
    \includegraphics[width=6.5in]{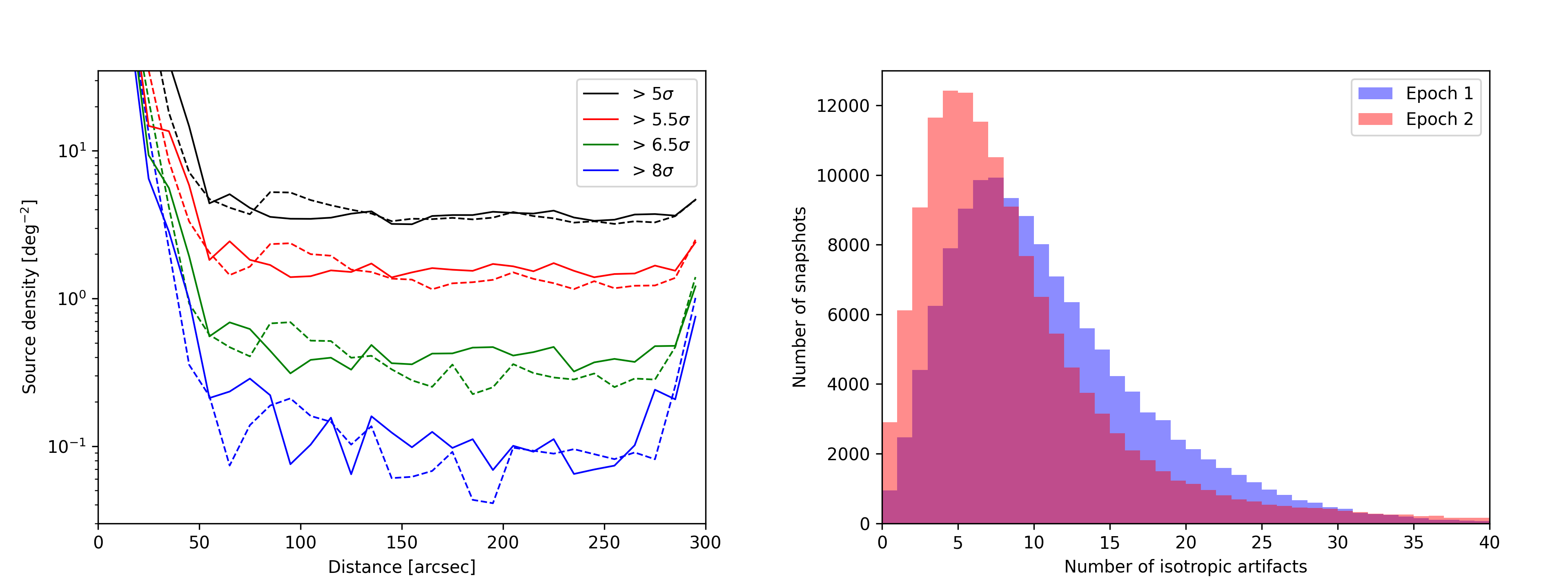}
    \caption{{\it Left:} Average source density around isolated point sources in VCSS epoch 1 ({\it dashed}) and epoch 2 ({\it solid}). Densities above several SNR cuts are shown. {\it Right:} Histograms of the number of isotropic artifacts in VCSS snapshots.}
    \label{artifacts}
\end{figure}

The left panel of Figure~\ref{artifacts} shows the stacked VCSS source density binned in annuli of $10^{\prime\prime}$ width. The inner two bins are dominated by detections of the IPS. Beyond $20^{\prime\prime}$ a distinct population of inner artifacts exist that dominates within $\sim 60^{\prime\prime}$ but has rapidly declining density with distance. A second artifact population dominates $> 60^{\prime\prime}$ that is isotropic with approximately constant density and similar amplitude across epochs. The density is SNR-dependent with declining numbers at higher SNR. The density upturn in outer annuli is due to the inner artifact population around sources just outside the $300^{\prime\prime}$ isolation radius.

From these results we conclude VCSS artifacts can be identified as sources without TGSS or RACS catalog matches and that the nearest neighbor distance can be used to determine which population artifacts belong to. We use this method to count the total number of isotropic artifacts with $SNR > 5$ in each snapshot.

The right panel of Figure~\ref{artifacts} shows the isotropic artifact count distribution in snapshots within either the TGSS or RACS footprint. Over $99.7\%$ of snapshots in each epoch are within this footprint. There is a large variation in the number of artifacts per snapshot with distribution peaks at $5-10$ artifacts. 

\section{False Association Simulations} \label{app:sim}

\begin{figure}[ht]
    \centering
    \includegraphics[width=6.5in]{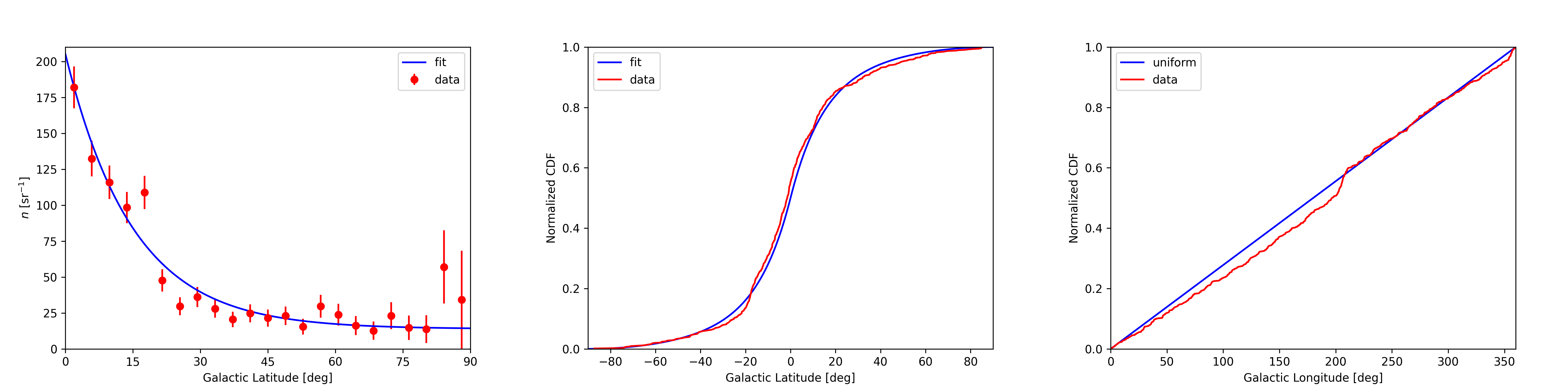}
    \caption{{\it Left}: Surface density with galactic latitude of the hot magnetic star catalog and the fitted equation~(\ref{eqn:fit}). {\it Middle}: Normalized galactic latitude CDF for the catalog and fit. {\it Right}: Normalized galactic longitude CDF for the catalog and a continuous uniform distribution.}
    \label{pdfcdf}
\end{figure}

To confirm our statistical calculations we ran simulations generating $50,000$ random catalogs using a model for the observed sky distribution. We assume the distribution is symmetric about the Galactic Plane and plot the stellar density binned in galactic latitude in the left panel of Figure~\ref{pdfcdf}. We fit the density with an exponential function:
\begin{equation}
\label{eqn:fit}
    n = n_0 \: e^{-k |b|} + c
\end{equation}
where $n$ is the star density per steradian,  $b$ is the galactic latitude in radians, and $n_0$, $k$ and $c$ are constants. The fitted values and $1\sigma$ uncertainties are ($n_0,k,c$) = ($191 \pm 12$, $3.8 \pm 0.5$, $13 \pm 4$). The two bins at highest declination were not used in fitting due to large uncertainties from the paucity of sources.

The middle panel of Figure~\ref{pdfcdf} compares the catalog normalized galactic latitude cumulative density function (CDF) to a numerical integration of the fitted stellar density function. The fit is a good approximation to the data. The right panel of Figure~\ref{pdfcdf} shows the CDF of a uniform distribution is a fair approximation to the catalog distribution in galactic longitude.

\begin{figure}[ht]
    \centering
    \includegraphics[width=6.5in]{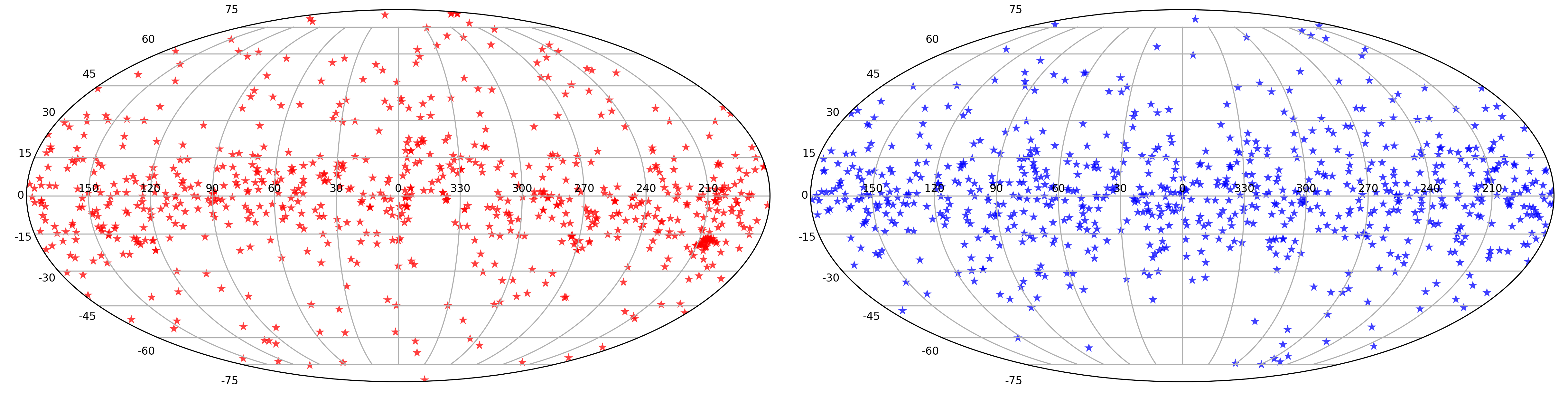}
    \caption{{\it Left}: Catalog of 761 hot magnetic stars plotted in Galactic coordinates. {\it Right}: A randomly generated catalog with the same number of stars and similar distribution.}
    \label{sky}
\end{figure}

Mock catalogs are constructed using an inverse transform technique to generate random coordinates for 761 stars following the model distribution. A sample from a random variable uniformly distributed between 0 and 1 is generated. The star's galactic latitude is set to where the CDF equals this number. Another uniform random variate is used to set the galactic longitude assuming stars are uniformly distributed along this axis. Figure~\ref{sky} compares the distributions of the stellar catalog and a mock catalog.

\begin{figure}[htb]
    \centering
    \includegraphics[width=3in]{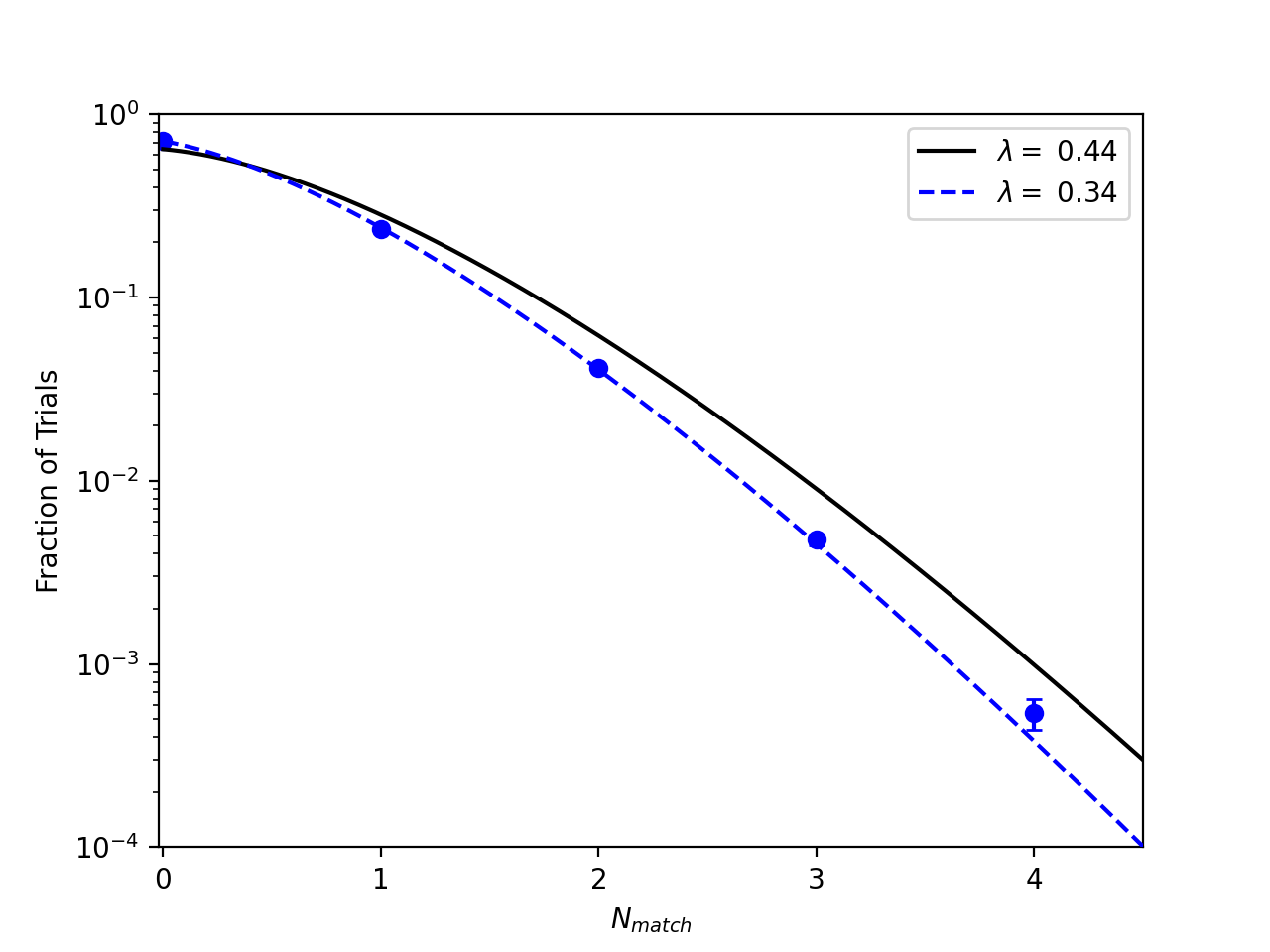}
    \caption{Fraction of 50,000 simulated catalogs having $N_{match}$ matches to the VCSS catalog. A Poisson distribution with expected value equal to the average number of matches (0.34) is a good fit to the simulations (dashed line). The simulations agree with the calculations for the stellar catalog (solid line) that the probability of three VCSS matches is $< 1\%$.}
    \label{trials}
\end{figure}

The number of VCSS matches is calculated for each mock catalog. The fraction of catalogs having a given number of matches is shown by the blue points in Figure~\ref{trials}. The  distribution is well described by Poisson statistics with an expected value equal to the average number of matches, $\lambda = 0.34$ (dashed line). Only 238 mock catalogs have three VCSS matches, giving a false association probability of $0.5\%$. This is lower than our calculation for the stellar catalog (solid line) and likely due to assumptions made on the stellar distribution: symmetry about the Galactic Plane and smoothly distributed in longitude (no clumps); as well as uncertainty in our fit to the density function. The simulations agree, however, that the probability of all three VCSS hot magnetic star detections are due to false associations with artifacts is $< 1\%$.

\bibliography{hms}
\bibliographystyle{aasjournal}



\end{document}